\documentclass[reprint,amsmath,amssymb,aps,prb,nofootinbib]{revtex4-2}
\usepackage{times}
\usepackage{subcaption}
\usepackage{graphicx}
\usepackage[dvipsnames]{xcolor}
\usepackage[font=small,
   justification=Justified,
   format=plain]{caption}
\usepackage{dcolumn}
\usepackage{physics}
\usepackage{siunitx}
\usepackage{dsfont}
\usepackage{bm}
\usepackage{hyperref}
\usepackage[mathlines]{lineno}
\usepackage{xcolor}
\usepackage{ulem}
\usepackage{comment}

\begin{document}

\title{Unconventional early-time relaxation in the Rydberg chain}

\author{Martin Schnee}
 \email{martin.schnee@usherbrooke.ca}
\author{Roya Radgohar}
\author{Stefanos Kourtis}
\affiliation{Institut quantique \& Département de physique, Université de Sherbrooke, Sherbrooke, Québec, Canada, J1K 2R1}

\date{\today}

\begin{abstract}
We show that unconventional relaxation dynamics of special initial states in one-dimensional arrays of Rydberg atoms produce non-generic decay of the initial-state survival probability (SP) at very early times. Using the PXP hamiltonian as a minimal model of the Rydberg blockade, we prove that the early-time SP for states exhibiting quantum many-body scarring (QMBS) decays at a characteristic rate, whose finite-size scaling is determined solely by scars. We numerically investigate the effects of both revival-enhancing and ergodicity-restoring deformations of the PXP hamiltonian and find results consistent with the limiting cases of integrable and ergodic dynamics, respectively. We moreover argue that such unconventional early relaxation of scarred initial states is characteristic of a whole class of QMBS models. Since the SP can be easily measured experimentally, our findings enable us to probe the presence of scars at time scales much shorter than that of thermalization.
\end{abstract}

\maketitle

\section{Introduction \label{sec:intro}}

Since its discovery in a Rydberg atom array~\cite{Bernien2017}, weak ergodicity breaking in quantum systems with constrained dynamics is under active investigation~\cite{Zhao2020, Bluvstein2021, Chen2022, Su2023, Gustafson2023}. This phenomenon has been associated with quantum many-body scars (QMBSs), special non-thermal mid-spectrum eigenstates. They have been shown to be present in varying number in a wide range of models, going from $O(1)$ to a number polynomial in the system size.

So far, experimental observations of QMBS have been achieved only for models with multiple scars. In this case their presence can be detected by quenching a fine-tuned initial state having high overlap with the QMBS and look for oscillations that persist for long times, much longer than the expected thermalization time of the system. However, this is challenging in quantum simulators with short coherence windows. Moreover, certain scarred dynamics do not lead to coherent revivals, an obvious example being models with a single scar. Uncovering signatures of QMBSs accessible at much earlier times and for a broader range of models would therefore render weak ergodicity breaking and associated effects more amenable to noisy quantum simulation.

In this paper, we trace the presence of QMBSs in the early-time decay of survival probability (SP) of initial states leading to scarring. The SP~\footnote{Also called revival fidelity, nondecay probability, or return probability.} is defined as the overlap between an initial state $|\psi_0\rangle$ and the evolved state $|\psi_t\rangle$ at time $t$:
\begin{equation}
    \mathcal{S}_\text{P} (t) = \big|\langle \psi_0 | \psi_t \rangle\big|^2 \,.
\end{equation}
At short times, the SP displays a universal quadratic decay~\cite{Flambaum2001}: 
\begin{equation}
    \mathcal{S}_\text{P}(t) \approx 1 - \sigma^2 t^2 \,.\label{eq:sp}
\end{equation}
The decay rate $\sigma^2$ is characteristic of the underlying dynamics and the initial state, and provides global information about the stability of quantum dynamics in isolated interacting quantum systems~\cite{TorresHerrera2014_1, TorresHerrera2014_2}. For evolution under a hamiltonian $H$, $\sigma^2$ is determined by the spread of the initial state over the energy eigenbasis $\{ |E_\alpha\rangle \}$ of $H$ as
\begin{align} 
    \sigma^2 &= \langle H^2 \rangle_{\psi_0} - \langle H \rangle_{\psi_0}^2 \label{eq:variance}\\
    &= \sum_{\alpha} \big|C_{\alpha}^0\big|^2 \, (E_\alpha - E_0)^2 \,,
\end{align}
where $\langle . \rangle_{\psi_0}$ denotes expectation with respect to $| \psi_0 \rangle$, $E_0 = \langle H \rangle_{\psi_0}$ is the energy of the initial state, and $C_{\alpha}^0 =  \langle\psi_0|E_\alpha\rangle$. In other words, the decay rate $\sigma^2$ is the variance of the overlap distribution $P_{\alpha}^0 = \big|C_{\alpha}^0\big|^2$ of the initial state over energy eigenstates. In the thermodynamic limit, the discrete probability distribution $P_{\alpha}^0$ becomes $\mathcal{P}{\scriptscriptstyle\text{LDOS}}(E) = \sum_{\alpha} |C_{\alpha}^{(0)}|^2 \;\delta(E-E_\alpha)$, known as the local density of states (LDOS). For simplicity, we use the term LDOS for both the discrete and the continuous case.

In generic quantum many-body systems with local interactions, initial states with energy close to the middle of the spectrum exhibit a Gaussian LDOS. It has been shown that their variance, and thus the SP decay rate, generically scales with system size $L$ as
\begin{equation}
    \sigma_{\text{gen.}}^2 \propto J^2 L
\end{equation}
where $J$ is the main energy scale of the system~\cite{TorresHerrera2014_1, TorresHerrera2014_2, Schiulaz2019}.

\begin{figure*}[t]
    \begin{subfigure}[b]{0.32\textwidth}%
    \includegraphics[width=\textwidth]{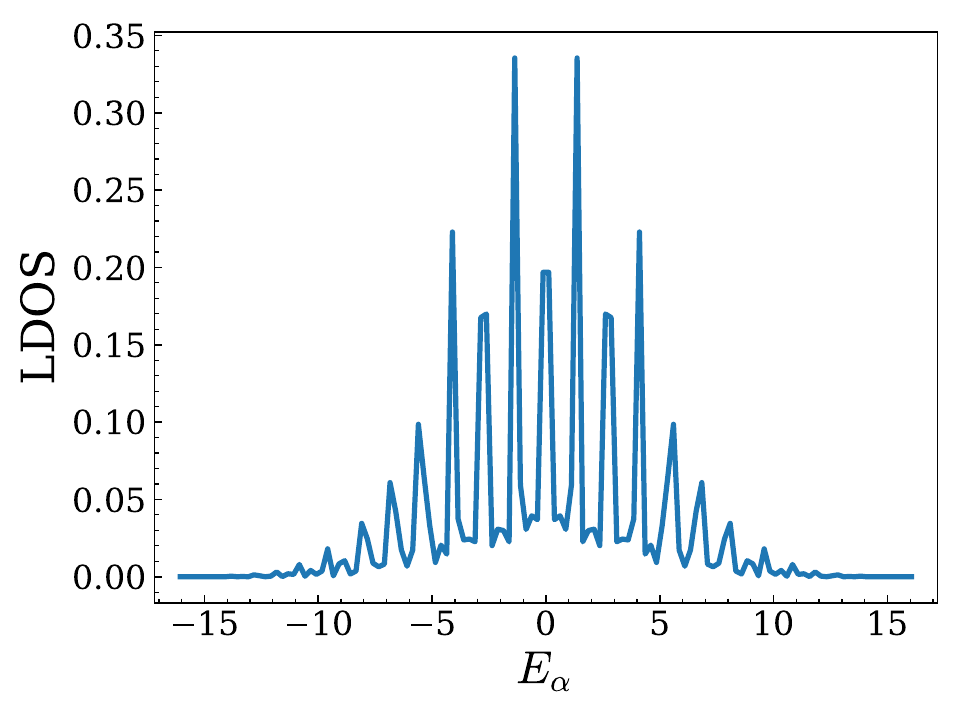}
    \caption{\label{fig:LDOSpxpXZ}}
    \end{subfigure}
\hfill%
    \begin{subfigure}[b]{0.32\textwidth}%
    \includegraphics[width=\textwidth]{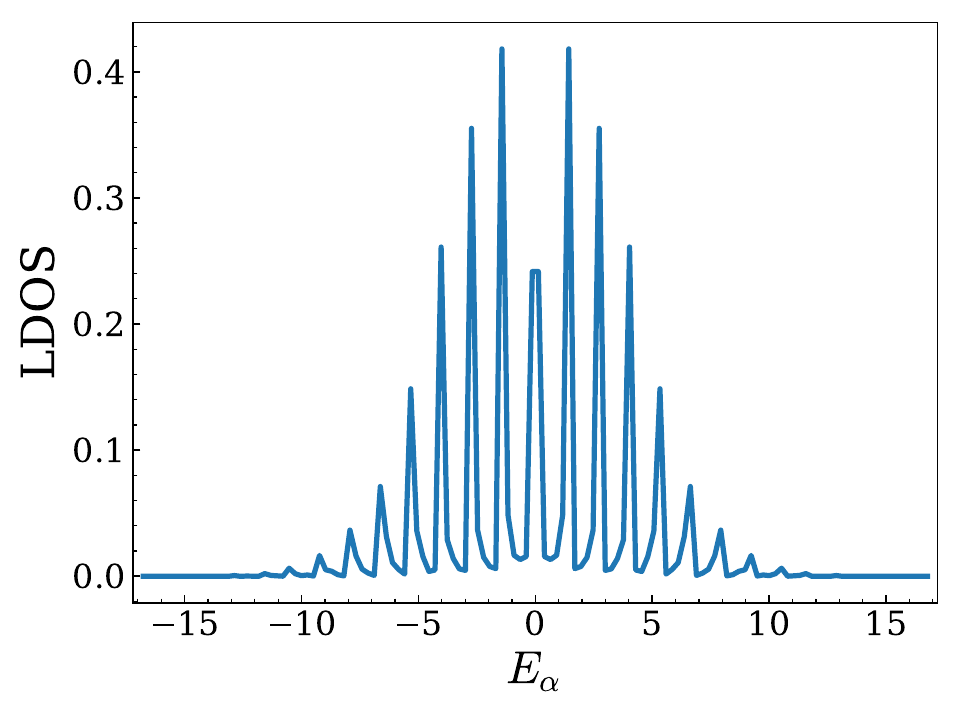}
    \caption{\label{fig:LDOSpxp}}
    \end{subfigure}
\hfill%
    \begin{subfigure}[b]{0.32\textwidth}%
    \includegraphics[width=\textwidth]{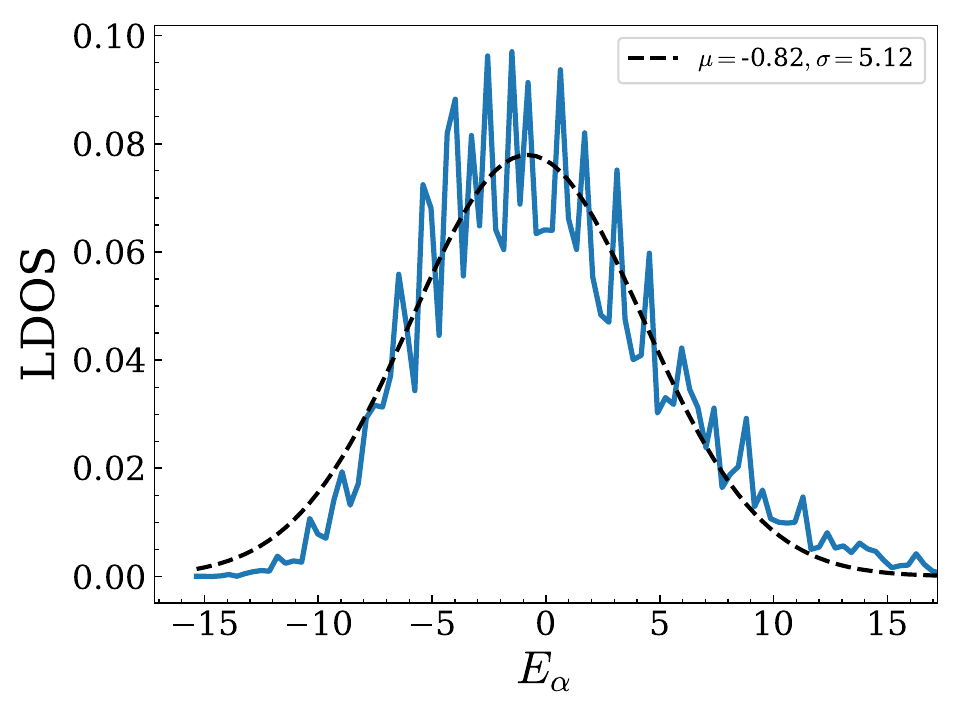}
    \caption{\label{fig:LDOSpxpNNN}}
    \end{subfigure}
\caption{LDOS of the $\ket{\mathbb{Z}_2}$ state over the energy eigenbasis of PXP model with $J=1$ (\subref{fig:LDOSpxpXZ}) with revival-enhancing deformation ($g_{\text{PXPZ}}=0.048$), (\subref{fig:LDOSpxp}) without deformation, and (\subref{fig:LDOSpxpNNN}) with ergodicity-restoring deformation ($g_{\text{NNN}}=1$). The black dashed line is a Gaussian fit for the LDOS. All plots are obtained by Lanczos exact diagonalization for $L=28$ qubits with 400 iterations (histogrammed with 130 bins).}
\end{figure*}

In contrast, we argue below that the LDOS of scarred initial states generally diverges substancially from this Gaussian form, causing a non-generic SP decay that can be seen at times significantly shorter than those usually considered to probe scarring effects. Beyond perfect scars, we show it holds in a paradigmatic model of approximate scars, the PXP model, which is also of experimental relevance due to its proximity with Rydberg-atom experiments. In both cases, the intuition is that the LDOS of initial states leading to scarred dynamics is dominated by a comb of QMBS contributions (see, e.g., Fig.~\ref{fig:LDOSpxp}), which fix $\sigma^2$. Assuming a semi-phenomenological model of a scarred LDOS, we derive general conditions under which the presence of QMBSs in the energy spectrum can be probed by measuring the early-time decay of the SP. Finally, we broaden the scope of our study by also distinguishing the cases of a single scar and of a few scars. The SP initial decay thus provides a sought-after early-time signature of QMBSs accessible in current rapidly-decohering quantum simulators.


\section{QMBS models \label{sec:model}}

There exist many recipes to construct nonintegrable models having QMBS in their spectrum. They can be divided in three main classes \cite{Serbyn2021}: models built from spectrum-generating algebra (SGA), models with Krylov restricted thermalization and models built by projector-embedding. 

These classes may exhibit specific QMBS features relevant for the behavior of the early-time SP decay. In the SGA class QMBS are equidistant and in number generally polynomial in $L$, in Krylov restricted models QMBS may or may not be equidistant and the number of QMBS can be polynomial or exponential in $L$. Projector-embedding models in the literature generally feature $\mathcal{O}(1)$ scars. We focus on example models for which simple scarred initial states are known.

\subsection{Perfect QMBS models}

We consider models in the SGA class to illustrate the case of initial states having overlap with QMBS equidistant in energy. In the SGA framework \cite{Mark_2020_1, Mark_2020_2, Moudgalya_2020}, QMBS eigenstates are constructed by repeatedly applying an operator $Q^+$ on a source state $|\psi_0 \rangle$ until the resulting state vanishes. This generates a tower of states: $\left\{|\mathcal{S}_k \rangle \right\}_{k=0,\dots,n} = \left\{ |\psi_0 \rangle \right.$, $Q^+|\psi_0 \rangle$, $(Q^+)^2|\psi_0 \rangle$,..., $\left. (Q^+)^n|\psi_0 \rangle \right\}$. Then, provided a given nonintegrable hamiltonian $H$ satisfies $[H, Q^+] |\psi_0 \rangle = \mathcal{E} Q^+ |\psi_0 \rangle$, and $|\psi_0 \rangle$ is an eigenstate of $H$ with energy $E_0$, then all the states in the tower will be eigenstates of $H$ equally spaced by an energy gap $\mathcal{E}$, i.e., $H |\mathcal{S}_k \rangle = (E_0 + k \mathcal{E}) |\mathcal{S}_k \rangle$. Provided the source state $|\psi_0\rangle$ is sufficiently simple, and $Q^+$ is a local operator, then its application does not increase the entanglement too much and the $|\mathcal{S}_k \rangle$ remain sufficiently non-thermal. Since $H$ was taken to be nonintegrable we are guaranteed to have embedded non-thermal eigenstates in an otherwise thermal spectrum. 

One simple way to construct models satisfying this constraint leads to hamiltonians with the following form \cite{Chandran_2023}:
\begin{equation}
    H = J \ H' + h \ Q^z \,.
    \label{eq:sga_ham}
\end{equation}
Here, the $Q^z$ term is a generator (or a linear combination of generators) of a nonabelian symmetry $G$, for which $Q^+$ is a raising operator such that $[Q^z, Q^+] = + \omega Q^+$ ($\omega$ being fixed). Thus the scar sector is associated to the energy scale $h$, and the energy spacing between scars will be $\mathcal{E}=\omega h$. The source state $|\psi_0 \rangle$ is chosen to be the lowest weight eigenstate of a unique irreducible representation of the group $G$. The term $H'$ is constructed so that the states in the tower generated by $Q^+$ are eigenstates of $H$ while ensuring that only the scar subspace is $G$-symmetric. This term is associated to a different energy scale $J$.

The literature provides many instances of such models for which simple initial states are known that are superpositions of the QMBS only \cite{Mark_2020_1, Mark_2020_2, Schecter_2019, Iadecola_2020}, i.e., $|\psi \rangle = \sum_{k=0}^{n} c_k |\mathcal{S}_k \rangle$. These scarred initial states are disconnected from the thermal part of the spectrum and lead to revivals, a distinctive feature of scar phenomenology, due to the equal spacing in energy of the $|\mathcal{S}_k \rangle$.

As a concrete representative of the SGA class, we consider the spin-1 XY model
\begin{align}
    H &= J \ H_{\text{XY}} + h \ (2J^z)\\ 
    &= J \ \sum_{i=1}^{L} (S_i^x S_{i+1}^x + S_i^y S_{i+1}^y) + h \ \sum_{i=1}^{L} S_i^z \,,
\end{align}
where the local spin-1 basis is $|+\rangle$, $|0\rangle$, $|-\rangle$, corresponding to eigenvalues $+1$, $0$, $-1$ of the local $S_z$ operator. This model was shown to exhibit a tower of $L+1$ exact scar eigenstates \cite{Schecter_2019}, $| \mathcal{S}_n \rangle = \mathcal{N}(n) \big(J^+ \big)^n | - \rangle^{\otimes L}$, with equally spaced energies $E_n = h(2n-L)$, where $n=0,\dots,L$ and $\mathcal{N}(n) = \sqrt{\frac{(L-n)!}{n!L!}}$ is a normalization coefficient. The scars obey an SU(2) algebra with the raising operator taking the form $J^+ = \frac{1}{2} \sum_{j=1}^{L} e^{i \pi j} \left( S_j^+ \right)^2$.

\subsection{Approximate QMBS in the Rydberg chain}

Approximate QMBS are known to arise in 1D chains of $L$ interacting Rydberg atoms (qubits) subjected to Rydberg blockade \cite{Bernien2017}. The Rydberg-atom hamiltonian describing this system is
\begin{equation}
    H_{\text{Ryd}} =  \frac{\Omega}{2} \sum_i X_i - \Delta \sum_i Q_i + \sum_{i} V_{i, i+1} Q_i Q_{i+1} \,,
    \label{eq:rydberg}
\end{equation}
where $\Omega$ is the Rabi frequency ($\hbar = 1$), $\Delta$ is the detuning parameter, $V_{i,j} \propto 1/|i-j|^6$ is the van der Waals interaction between the atoms. Here,  $Q_i =  (1+Z_i)/2 = \dyad{0}$ count local Rydberg excitations and $X_i$, $Z_i$ are Pauli matrices associated to the $i$-th qubit. In the strongly interacting Rydberg blockade regime, which corresponds to neglecting interactions beyond nearest neighbors (i.e. $V_{i,j}$ for $j \geq  i+2$), and taking the limit $V_{i, i+1} \gg \Omega$, the low-energy subspace can be described by the effective PXP model (at $\Delta=0$):
\begin{equation}
    H_{\text{PXP}} = J \sum_{i=1}^L P_{i-1} X_i P_{i+1} \,,
    \label{eq:pxp_ham}
\end{equation}
where $J=\frac{\Omega}{2}$, and projectors $P_i = (1-Z_i)/2 = \dyad{1}$ enforce the Rydberg blockade constraint. Hereafter, we consider periodic boundary conditions.

A quench of the state $\ket{\mathbb{Z}_2} = \ket{1010\dots}$ with the PXP hamiltonian leads to long-lived revivals of the initial state. This behavior can be traced back to the significant but approximate overlap of the $\ket{\mathbb{Z}_2}$ state with $k=L-1$ QMBS eigenstates. These QMBS are not exactly equally spaced in energy and no exact form is known for most of them. This indicates the existence of an approximate SU(2) algebra emerging from the $\ket{\mathbb{Z}_2}$ state, suggesting an approximate SGA~\cite{Serbyn2021}, however without any distinctive energy scale for the scar sector.

\section{Non-generic decay of perfectly scarred initial states \label{sec:analytics}}


We show that the early-time SP decay rate of initial states that are superposition of exact scars, $|\psi \rangle = \sum_{k=0}^{n} c_k |\mathcal{S}_k \rangle$, will distinctively diverge from the generic decay since it is controlled by scar-specific features: (i) the energy scale of the QMBS sector $h$, and (ii) $|c_k|^2$ the scar overlap shape for the relevant initial state. 

The survival amplitude is indeed given by: $\langle \psi| e^{-iHt} |\psi \rangle = e^{iE_0 t} \sum_{k=0}^{n} |c_k|^2  e^{-ikht}$. Taking the square modulus leaves only a periodic dependence in the energy scale of the scar sector $h$ with period $2\pi/h$ and on the shape of the overlap with scars eigenstates $|c_k|^2$. At short times ($t \ll \sigma^{-1}$), 
\begin{align}
    \mathcal{S}_P(t) &= \left| \langle \psi| e^{-iHt} |\psi \rangle \right|^2 \\&\approx 1 - h^2 \left(  \sum_{k=1}^{n} k^2 |c_k|^2 - \left(  \sum_{k=1}^{n} k |c_k|^2 \right)^2 \right) t^2 \,,
\end{align}
where the decay rate depends solely on the energy scale of the scar sector and the shape of the scar overlaps. In contrast, generic initial states will have overlap with both the scar sector and the thermal sector of $H$. The latter being exponentially larger than the former, the overlap distribution (LDOS) will be gaussian with a variance $\propto L$, controlled exclusively by the energy scale $J$ of the non-integrable part of $H$.  For the scarred initial states to exhibit the exact same decay as the generic ones the situation would therefore have to be fine-tuned. 

As a concrete example we consider a quench with the spin-1 XY model of the following product initial state:
\begin{align}
    |\psi \rangle &= \bigotimes_{j=1}^{L} \left( \frac{|+\rangle - (-1)^{j} |-\rangle}{\sqrt{2}} \right) \\&= \sum_{n=0}^{L} c_n \, | \mathcal{S}_n \rangle \ , \ \text{with } c_n = (-1)^n \sqrt{\frac{1}{2^L}\binom{L}{n}},
\end{align}
which is the ground state of the hamiltonian $H_0 = J^x$ and a superposition of perfect scars. Its SP can be computed exactly and depends only on scar features: $\mathcal{S}_P(t) = \cos(ht)^{2L}$. At short times ($t \ll \sigma^{-1}$), $\mathcal{S}_P(t) \approx 1 - \sigma^2 t^2$ with a decay rate 
\begin{equation}
    \sigma^2 = h^2 L    
\end{equation}
that depends on scars only. Additionnally, notice that this early decay time scale, $t \ll (\sqrt{L} h)^{-1}$, is much smaller than the time scale of the first revival, $t_{\text{rev}} = \pi/h$, and the difference increases as $L$ grows.

On the contrary, a generic initial state will have overlap on the exponentially larger thermal sector. This sector is associated to the nonintegrable part $H'$ of equation \ref{eq:sga_ham}. Thus the decay rate will be associated to the energy scale $J$. For example, for the spin-1 XY model, the generic initial state $|\psi \rangle = | 0 \rangle^{\otimes L}$ is in the middle of the spectrum (zero mean energy), and leads to a decay rate 
\begin{equation}
    \sigma^2 = J^2 L .   
\end{equation}
This spin-1 XY model with QMBS can be generalized to a non-translation invariant model. It can also be generalized to higher spatial dimension $d$, displaying QMBS still equidistant in energy but whose number scales with $L^d$. The above SP behavior for the initial state $|\psi \rangle$ can also be extended to higher $d$ replacing $L$ by $V=L^d$ in the formulae, so that $\mathcal{S}_P(t) = \cos(ht)^{2V}$ and the  early-time decay rate is $\sigma^2 = h^2 V$.

\section{Numerical results for approximate scars \label{sec:numerics}}

In this section, we consider the $\ket{\mathbb{Z}_2}$-state dynamics of the PXP model as a paradigmatic example of approximate scarring and numerically demonstrate that the resulting non-generic early relaxation dynamics is solely controlled by scars as well.

The $\ket{\mathbb{Z}_2}$ LDOS takes the form of $L-1$ narrow peaks aligned with QMBS energies, approximately equally separated by a size-independent energy gap~\cite{Turner2018, Turner2018_2} --- see Fig.~\ref{fig:LDOSpxp}. The revival period is inversely proportional to this gap. The resulting LDOS differs significantly from the usual Gaussian distribution observed in generic evolution (see, e.g., Fig.~\ref{fig:LDOSpxpNNN}).

The variance of the full LDOS can be analytically calculated from Eq.~\eqref{eq:variance}, and is given by $\sigma_{\text{PXP} | \mathbb{Z}_2}^{2} = J^2 \frac{L}{2}$, since $\langle H_{\mathrm{PXP}} \rangle_{\mathbb{Z}_2} = 0$ and only $L/2$ terms contribute to the expectation $\langle H_{\mathrm{PXP}}^2 \rangle_{\mathbb{Z}_2}$. This non-generic $L/2$ scaling distinguishes the scarred LDOS from the generic Gaussian case~\cite{TorresHerrera2014_1, TorresHerrera2014_2, Schiulaz2019}. In contrast, an analogous calculation for the rapidly thermalizing initial state $\ket{\mathbb{Z}_1} = \ket{1}^{\otimes L}$ yields the generic behaviour $\sigma_{\text{PXP} | \mathbb{Z}_1}^{2} = J^2 L$.

We stress that the corresponding time scales allowing for distinguishing between the scarred and the non-scarred behavior are much smaller than the first revival time scale. For concreteness, for realistic experimental parameters we find the characteristic early SP decay is observable at timescales $T_{\text{PXP} | \mathbb{Z}2} \approx 0.16 \mu s$. This is more than $30$ times smaller than both the natural timescale for local relaxation, $T_{\text{therm.}}$, and the timescale of the first initial state revival, $T_{\text{rev}}$, the separation with the latter timescale increasing as $\sqrt{L}$.
Our results show that the SP decay can be used as an experimental probe of scarring in the Rydberg blockade in parameter regimes for which the revival time scale would be prohibitively long. 
We present a more thorough discussion of experimental time scales in Sec.~\ref{sec:discussion}.

\begin{figure}
    \hspace*{-0.3cm}
    \includegraphics[width=0.5\textwidth]{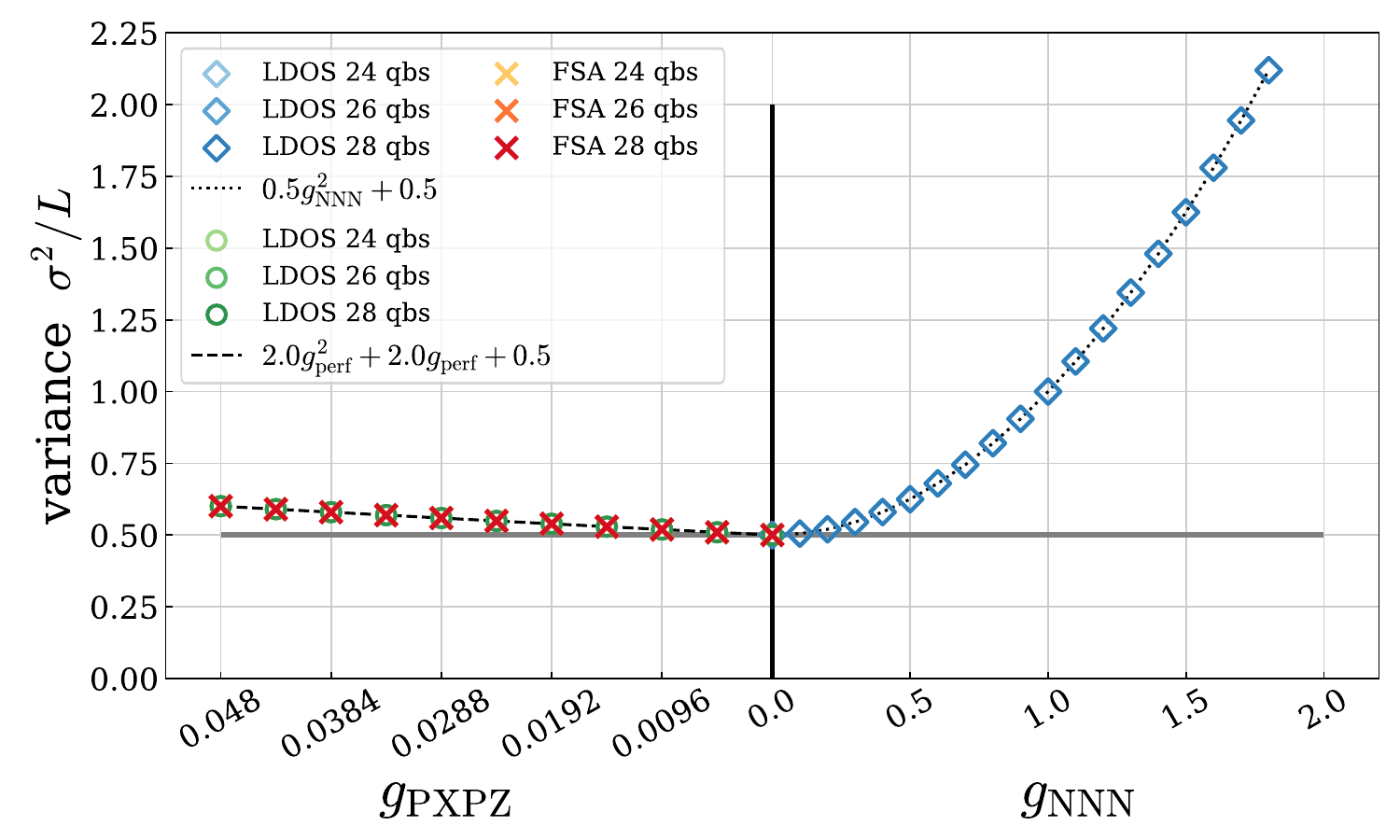}%
    \caption{FSA scars variance vs LDOS variance rescaled with system size as a function of deformation strength, for $L=24,\, 26,\, 28$ sites and $J=1$. (Right half) Variance when PXP is strongly deformed toward restoration of ergodicity. (Left half) Variance when PXP is perturbed toward perfect scarring behavior. The bare PXP model is at $g_{\text{PXPZ}}=g_{\text{NNN}}=0$. Black dashed lines are polynomial fit for a fixed system size $L=28$. Full LDOS data are obtained from regular Lanczos method with 400 iterations. \label{fig:varscalingallpert}}
\end{figure}

We now show that the non-generic scaling of $\sigma^2$ is determined exclusively by QMBSs in the PXP model, thus signalling their presence. An excellent approximation of QMBS eigenstates can be obtained using the so-called forward-scattering approximation (FSA)~\cite{Turner2018_2,Nandy_2024}. The FSA allows us to numerically isolate QMBS contributions to the LDOS. The FSA calculation yields a variance $\sigma_{\text{PXP} | \mathbb{Z}_2}^{2 \,\text{(FSA)}} = L/2$ --- see Fig.~\ref{fig:varscalingallpert} (we take $J=1$ for all the numerics).

To further reinforce the conclusion that scars alone, if present, determine the early-time SP in the PXP model, we study the effect of deformations $\delta H$ that either enhance revivals or restore ergodicity. In what follows, we only consider deformations and initial states for which $\langle \delta H \rangle_{\psi_0} = 0$. Thus, upon introducing such a deformation, the LDOS variance for the total hamiltonian $H = H_{\text{PXP}} + \delta H$ becomes
\begin{equation}
    \sigma^2 = \langle H_{\text{PXP}}^2 \rangle_{\psi_0} + \langle H_{\text{PXP}} \delta H \rangle_{\psi_0} + \langle \delta H H_{\text{PXP}} \rangle_{\psi_0} + \langle \delta H^2 \rangle_{\psi_0} \,.
    \label{eq:mvsquared}
\end{equation}
Below we evaluate this expression for $\ket{\psi_0} = \ket{\mathbb{Z}_2}, \ket{\mathbb{Z}_1}$ and two different deformations. The effect of these perturbations on the $\ket{\mathbb{Z}_2}$ LDOS are shown in Fig.~\ref{fig:LDOSpxpXZ},\ref{fig:LDOSpxpNNN}.

We first introduce a term known to enhance the fidelity of the revivals in the perturbative regime~\cite{Khemani2019,Choi2019}. This reads
\begin{equation}
H_{\text{PXPZ}} = g_{\text{PXPZ}} \sum_j P_{j-1} X_j P_{j+1}(Z_{j+2} + Z_{j-2}) \,,
\end{equation}
with $g_{\text{PXPZ}} \geq 0$. Its effect is to increase the rate of flips on qubit $j$ when a qubit is in $|0\rangle$ two sites away. Nearly perfect revivals are obtained for $g_{\text{PXPZ}}\approx 0.048$. 

The $\ket{\mathbb{Z}_2}$ LDOS with respect to the total hamiltonian $H$ with $\delta H = H_{\text{PXPZ}}$ remains highly peaked within the range $g_{\text{PXPZ}} \in [0, 0.048]$, with the number of peaks staying constant. For the $\ket{\mathbb{Z}_2}$ initial state, the terms in Eq.~\eqref{eq:mvsquared} are evaluated as $\langle H_{\text{PXP}} H_{\text{PXPZ}} \rangle_{\mathbb{Z}_2} = \langle H_{\text{PXPZ}} H_{\text{PXP}} \rangle_{\mathbb{Z}_2} = J g_{\text{PXPZ}} L$ and $\langle H_{\text{PXPZ}}^2 \rangle_{\mathbb{Z}_2} = 2 g_{\text{PXPZ}}^2 L$. The LDOS variance is therefore $\sigma_{\text{PXPZ} | \mathbb{Z}_2}^{2} = (4g_{\text{PXPZ}}^2 + 4 J g_{\text{PXPZ}} + J^2) \frac{L}{2}$. The scar variance extracted from FSA numerics agrees perfectly with this scaling, as shown in Fig.~\ref{fig:varscalingallpert}. In contrast, the LDOS variance of the $\ket{\mathbb{Z}_1}$ state remains Gaussian without dominant overlap with scars. The scaling of its variance is distinct from that of $\ket{\mathbb{Z}_2}$. We namely find $\sigma_{\text{PXPZ} | \mathbb{Z}_1}^{2} = (4g_{\text{PXPZ}}^2 - 4 J g_{\text{PXPZ}} + J^2 ) L$. In the range where revival enhancement is observed for the $\ket{\mathbb{Z}_2}$ initial state, $g_{\text{PXPZ}} \ll 1$, we notice that $H_{\text{PXPZ}}$ has an \textit{opposite} effect on the LDOS variance of the $\ket{\mathbb{Z}_2}$ and $\ket{\mathbb{Z}_1}$ initial states.

Next, we deform the PXP hamiltonian with a term that destroys the special eigenstates and their associated coherent revivals, as well as any trace of “slow” thermalization in eigenstates, and restores strong eigenstate thermalization~\cite{Turner2018_2}. This deformation describes next-nearest-neighbor correlated flips:
\begin{equation}
    H_{\text{NNN}} = g_{\text{NNN}} \sum_j P_{j-1} X_j P_{j+1} X_{j+2} P_{j+3} \,.\label{eq:HNNN}
\end{equation}
As can be seen in Fig.~\ref{fig:LDOSpxpNNN}, $H_{\text{NNN}}$ suppresses the peaks in the $\ket{\mathbb{Z}_2}$ LDOS and a generic Gaussian LDOS is observed for large enough $g_{\text{NNN}}$. We note that restoration of ergodicity means that the FSA is no longer valid, even for small $g_{\text{NNN}}$. The exact expression for the variance scaling is $\sigma_{\text{NNN} | \mathbb{Z}_2}^{2} = (g_{\text{NNN}}^2 + J^2) \frac{L}{2}$. The $\ket{\mathbb{Z}_1}$ LDOS behaves similarly in this case, with scaling $\sigma_{\text{NNN} | \mathbb{Z}_1}^{2} = (g_{\text{NNN}}^2 + J^2 ) L$. This is to be contrasted with the opposite trends obtained above for $\ket{\mathbb{Z}_2}$ and $\ket{\mathbb{Z}_1}$ initial states under the scar-enhancing deformation.

These results, summarized in Fig.~\ref{fig:varscalingallpert}, strongly suggest that, as soon as QMBS overlaps are dominant, they dictate the full LDOS variance and consequently the short-time SP decay. This picture is reminiscent of the exact behavior derived in the previous section for SGA-class QMBS models and perfectly scarred initial states. This is not surprising since the PXP model is known to belong approximately to the SGA class as it hosts only an approximate SU(2) symmetry \cite{Choi2019}. However, in the PXP model this feature is non-trivial to uncover. This is due to the scar sector being hybridized with the thermal sector, and the absence of distinct associated energy scales. Still, the scarred $\ket{\mathbb{Z}_2}$ state has a manifestly distinct decay than the generic state $\ket{\mathbb{Z}_1}$ state.

Because of the hybridization, the peaks in the $\ket{\mathbb{Z}_2}$ LDOS are of finite width, contrary to the perfect scar initial state considered in Sec.~\ref{sec:analytics} . The perfect agreement between the FSA scars decay and the real decay in our simulation shows this finite width has only negligible effects on the early decay. Fourier theory in fact tells us that if the energy scale of this peak width is small then it should translate into long time effects in the SP behavior. In particular, if the peak width is smaller than the energy gap between scars then time scale at which the finite peak width becomes significant will be beyond the first revival time scale. Additionally, if we assume a gaussian shape for the peaks, then the effect is an overall gaussian damping of the SP revivals, even slower as the width $\sigma_p$ is small. Thus, at short times and for large enough $L$, we expect the finite width of scars to be negligible. The perfect agreement between the FSA decay and the real decay shown in our work supports this claim.

To finish, notice that the PXP model has the advantage of being an approximate member of the other two classes of QMBS models as well, the projector embedding and the Krylov-restricted thermalization classes, demonstrating the fact that our detection scheme cares only about the number of scars and the magnitude of their overlap with the initial state, but is agnostic with respect to model details.

\section{Discussion \label{sec:discussion}}

\subsection{Conditions for scarred decay from a semiphenomenological model} 

Building on the insights collected so far, we shed light on the conditions under which the presence of QMBSs can be probed by measuring the early-time decay of the SP using a general model of a scarred LDOS. The semiphenomenological model that follows is based on the early idea that the energy spectrum of scarred models can be divided in a subspace of QMBS eigenstates and a subspace of thermal eigenstates~\cite{Serbyn2021}, $H \approx \mathcal{H}_{\text{scar}} \oplus \mathcal{H}_{\text{thermal}}$. The relevant initial states to probe the presence of QMBS will generally be those having an overlap over both sectors, but whose overlap with the QMBS sector is significantly higher than with the thermal one.

To motivate our scarred LDOS model, we consider the sum of the bare PXP hamiltonian, hosting approximate QMBSs, and any deformation $\delta \tilde H$ for which $\langle \delta \tilde H \rangle_{\psi_0} = \langle H_{\text{PXP}} \delta \tilde H \rangle_{\psi_0} = \langle \delta \tilde H H_{\text{PXP}} \rangle_{\psi_0} = 0$ for the initial state(s) of interest, such as the $H_{\text{NNN}}$ deformation introduced above. Due to the linearity of the expectation, this is the case whenever there are no terms in $\delta \tilde H$ that can undo the effect of the terms of $H_{\text{PXP}}$ on the initial state. Since each term of the latter contains an odd number of $X$ operators (one here), the local terms of the former must contain an odd number of them as well to cancel their effect. Therefore, for these deformations, the variance reduces to the sum of the variances of PXP and deformation terms: $\sigma^2 = \langle H_{\text{PXP}}^2 \rangle + \langle \delta\tilde H^2 \rangle$. Specifically, the effect of the ergodicity-restoring deformation $H_{\text{NNN}}$, whose terms are parity-preserving, can be seen as an independent contribution to the LDOS variance.

With the above reasoning in mind, we can abstract away all microscopic model details and focus on a simple model of scarred LDOS of the form
\begin{equation}
    \mathcal{P}_{\text{LDOS}} (E) = w_{\text{scar}} \, \mathcal{P}_{\text{scar}} (E) + w_{\text{th.}} \, \mathcal{P}_{\text{th.}} (E)
    \label{eq:toymodelLDOS}
\end{equation}
with the constraint $w_{\text{scar}} + w_{\text{th.}} = 1$ to ensure normalization. The distribution $\mathcal{P}_{\text{scar}}$ describes (approximate) scars and $\mathcal{P}_{\text{th.}}$ represents the background of thermal states, both normalized to 1.

Motivated by our results for the deformed PXP model discussed above, we assume that both distributions have the same mean, so that $\mu_{\text{LDOS}} = \mu_{\text{scar}} = \mu_{\text{th.}}$. However, we consider the general case for which $\sigma_{\text{scar}}^2 \neq \sigma_{\text{th.}}^2$, where $\sigma_{\text{scar}}^2$ and $\sigma_{\text{th.}}^2$ are the variances of the distributions of each sector. The total LDOS variance is then $\sigma_{\text{LDOS}}^2 = w_{\text{scar}} \sigma_{\text{scar}}^2 + w_{\text{th.}} \sigma_{\text{th.}}^2$. In this general setting, one finds that $\sigma_{\text{scar}}^2$ fully determines the observed $\sigma_{\text{LDOS}}^2$, i.e., $\sigma_{\text{LDOS}}^2 \simeq \sigma_{\text{scar}}^2$ if 
\begin{equation}
    \frac{w_{\text{scar}}}{w_{\text{th.}}}  \gg \frac{\sigma_{\text{th.}}^2}{\sigma_{\text{scar}}^2},
\end{equation}
meaning the weight ratio significantly “compensates” the inverse variance ratio between both scar and thermal sectors.

Moreover, we can consider the case where $\mathcal{P}_{\text{scar}}$ and $\mathcal{P}_{\text{th.}}$ in Eq.~\eqref{eq:toymodelLDOS} both have close but distinct extensive variances, $\sigma_{\text{scar}}^2 = \alpha L$ and $\sigma_{\text{th.}}^2 = \beta L$ (with $\alpha, \beta \in \mathcal{O}(1)$), as is the case for the PXP model with the ergodicity-restoring deformation~\eqref{eq:HNNN}. For the total variance to be determined by the scars variance up to an error $\epsilon$, i.e., $\left|\sigma_{\text{LDOS}}^2 - \sigma_{\text{scar}}^2 \right| \leq \epsilon$, we must have $w_{\text{th.}} \gamma L \leq \epsilon$. Since $\gamma$ is constant, requiring $\epsilon \rightarrow 0$ implies $w_{\text{th.}} \rightarrow 0$. In particular, requiring $\epsilon$ to be exponentially small in the system size implies the weight of the thermal background must vanish exponentially as well. Thus, this simple model of scarred LDOS allows one to make precise the intuition that QMBS contributions must be dominant in the LDOS to be able to probe their presence by experimental measurements of the SP at times much shorter than that of thermalization.

\subsection{Single and few scars} 

We now broaden the scope of applicability of the SP as an early-time signature of QMBSs by discussing the case of models with $\mathcal{O}(1)$ scars. Two particular situations stand out. In the first one, two scars are projector-embedded in the bulk of one-dimensional nonintegrable spin-$\frac{1}{2}$ chain such that their separation in energy is $\Delta E \propto L$~\cite{Ok_2019}. The early-time decay of an initial superposition of these two scars can be understood as resulting from an LDOS made of two gaussian peaks~\cite{TorresHerrera2014_3}. As long as the energy gap $\Delta E \propto 2L$ is significantly larger than the peak width, it defines the early-time decay rate. Moreover, the value of the gap is tuned by the specific choice of projector realizing the embedding.

The case of an initial state being a single QMBS eigenstate is trivial since the LDOS is a Dirac delta and the SP remains constant, making it impossible to distinguish from a generic volume-law entangled eigenstate. However,  the situation where the overlap with a single scar is only approximate, with contribution from eigenstates of nearby energies, is much more relevant. Then, the LDOS looks like a single narrow peak of finite width around the scar energy, giving a finite SP decay rate~\cite{Gorin_2006, Gorin_2016, TorresHerrera2014_3}. In particular, it was proposed in~\cite{Gotta_2023} to parametrically deform a single QMBS eigenstate and then take it as initial state. This initial state, while not scarred, keeps overlapping mainly with eigenstates whose energy is close to the original scar energy, and thus carries traces of the presence of the original scar that can be seen distinctively in the early SP decay, even in the thermodynamic limit. 

Essentially, a variant of the spin-1 XY hamiltonian is considered which retains the same set of scars, $H = J \ H_{\text{XY}} + h \ (2J^z) + D \sum_{i=1}^{L} (S_i^z)^2 + J_3 \sum_{\langle ij \rangle} (S_i^x S_{i+3}^x + S_i^y S_{i+3}^y)$. A single scar $| \mathcal{S}_n \rangle = \mathcal{N}(n) \ \big(J^+ \big)^n | - \rangle^{\otimes L} \equiv |n,\pi\rangle$ of the spin-1 XY model is selected and deformed by changing local phases, $|n,k\rangle = \tilde{\mathcal{N}}(n) \ J_k^+ \big(J^+ \big)^{n-1} | - \rangle^{\otimes L}$ with $J_k^+ = \frac{1}{2} \sum_{j=1}^{L} e^{i k j} \left( S_j^+ \right)^2$. The resulting state family, parameterized by $k$ ($k\neq \pi$ and integer multiple of $\frac{2\pi}{L}$), is not an eigenstate of the model anymore, however, (i) it has the same mean energy as the original QMBS, and (ii) it overlaps an exponential number of thermal eigenstates while remaining sub-volume-law entangled. 

Most importantly, the LDOS takes the form of a narrow gaussian peak with finite width. If one chooses a deformation parameter $k=\pi + \frac{2\pi}{L}m$ scaling as $1/L$ (with $m\in \mathbb{Z}$), so that the initial states $|n,k\rangle$ shift toward the exact scar eigenstate in the thermodynamic limit, then their energy variance over $H$ behaves as $\sigma^2 \sim (J^2 + 9J_3^2)(k-\pi)^2 \propto 1/L^2$.  This means the LDOS narrows and the SP decay slows down as the system size is  increasing, oppositely to the generic decay behavior. 

Let us recall that single scars cannot be detected from coherent revivals (at least two peaks in the LDOS are necessary for revivals). This fact has already motivated works the search for observable signatures~\cite{Larsen_2024} since no experimental observation of a single QMBS has been achieved to date. Looking at early SP decay of such single scar deformation may thus be the simplest way to detect the presence of single scars in relevant physical systems.

Notice that models in the projector-embedded class found in the literature often contain a single scar and thus may also fall in this category.

\subsection{Time scales and resources for experimental detection of SP decay}

Finally, we elaborate on the experimental parameters required to measure the scarred SP decay in the Rydberg chain effectively described by the PXP hamiltonian in Eq.~\eqref{eq:pxp_ham}. 

Quenching from the $\ket{\mathbb{Z}_2} = \ket{1010\dots}$ initial state, the SP decay is quadratic (see Eq.~\eqref{eq:sp}) with a rate given by $\sigma_{\mathbb{Z}_2}^2 = \left(\frac{\Omega}{2} \right)^2 \frac{L}{2}$ if $L$ even ($\left(\frac{\Omega}{2} \right)^2 \frac{L-1}{2}$ if $L$ odd, respectively). The timescale of interest in our case is $T_{\text{PXP} | \mathbb{Z}_2} \ll \sigma_{\mathbb{Z}_2}^{-1} \propto 1/\sqrt{L}$ and must be contrasted with the timescale of the first revival of the SP, $T_{\text{rev}} = 1.38 \times 2\pi/\Omega$, which is independent of $L$. This latter timescale is the minimum timescale current experiments have to reach to detect manifestations of QMBS with conventional methods. Moreover, the temporal separation between these two timescales is increasing as $\sqrt{L}$.

We now show, for a chain of length $L=9$ and a Rabi frequency $\Omega = 2\pi \times 0.2 \text{ MHz} \ll V$ (valid for typical interactions $V\gtrsim 1\text{ MHz}$ \cite{Browaeys_2020}), that this difference of timescales is significant even for small system sizes. We find the characteristic SP early decay is observable at timescales $T_{\text{PXP} | \mathbb{Z}_2} \ll \sigma_{\mathbb{Z}_2}^{-1}$, a reasonable estimate being $ T_{\text{PXP} | \mathbb{Z}_2} \leq \sigma_{\mathbb{Z}_2}^{-1}/5 = \SI{0.16}{\micro\second}$. This is more than $30$ times smaller than the natural timescale for local relaxation, $T_{\text{therm.}}=2\pi/\Omega \approx \SI{5.0}{\micro\second}$, and most importantly more than $40$ times smaller than the timescale of the first revival of the initial state, $T_{\text{rev}} \approx \SI{6.9}{\micro\second}$.

On the other hand, quenching from $|\mathbb{Z}_1\rangle = |111\dots\rangle$ the decay rate is given by $\sigma_{\mathbb{Z}_1}^2 = \left(\frac{\Omega}{2} \right)^2 L \ \text{, } \forall L$. Using the same parameter values we find the SP early decay is observable at timescales $T_{\text{PXP} | \mathbb{Z}_1} \ll \sigma_{\mathbb{Z}_1}^{-1}$, a reasonable estimate being $T_{\text{PXP} | \mathbb{Z}_1} \leq \sigma_{\mathbb{Z}_1}^{-1}/5 =  \SI{0.11}{\micro\second}$.

Can we distinguish the early decay behavior of $|\mathbb{Z}_1\rangle$ and $|\mathbb{Z}_2\rangle$? A necessary prior is to be able to make measurements at several times before this timescale. This rests upon having a fast measurement clock rate. On commercially available quantum simulators this clock rate is around $\SI{0.004}{\micro\second}$. Moreover, the shortest time from which measurements can be performed is $\sim \SI{0.050}{\micro\second}$. Considering the above timescales, this allows for 15 to 25 measurement points. Looking at slightly latter times by taking into account the following quartic term in the early-time SP decay gives time scale $T_{\text{PXP} | \mathbb{Z}_2}^{'} \leq w_{\mathbb{Z}_2}^{-1}/5 = \SI{0.5}{\micro\second}$, allowing for more measurement points while remaining at much shorter timescales than the first revival. 

The ability to distinguish the two SP decays in an experiment at these short timescales depends on the ability to estimate the SP with a small enough error bar. More precisely, the error bar $\epsilon$ must be smaller than the maximum difference between the two decays, $\delta_{\mathbb{Z}_2, \mathbb{Z}_1}= \mathcal{S}_P^{\mathbb{Z}_2}(t_{\text{max}}) - \mathcal{S}_P^{\mathbb{Z}_1}(t_{\text{max}})$, at least for the longest time considered. Apart from the imperfections of the experiment itself, the error $\epsilon$ depends on the number of experiments per point. For a single point at time $t$, the sample SP is given by $\frac{M_1}{M}$ with $M$ the total number of experiments performed and $M_1$ the number of times the $|\mathbb{Z}_i \rangle$ configuration is measured. Then the fluctuations of $\frac{M_1}{M}$ around the theoretical value $S_P (t)$ is proportional to $1/\sqrt{M}$ (shot noise). For $t_{\text{max}} = T_{\text{PXP} | \mathbb{Z}_1} = \sigma_{\mathbb{Z}_1}^{-1}/5$, this leads to $\delta_{\mathbb{Z}_2, \mathbb{Z}_1} = 1 - \sigma_{\mathbb{Z}_2}^2 (\sigma_{\mathbb{Z}_1}^{-1}/5)^2 - (1 - \sigma_{\mathbb{Z}_1}^2 (\sigma_{\mathbb{Z}_1}^{-1}/5)^2) = \frac{3}{50}$ if $L$ even ($\frac{1}{50} \frac{3L-1}{L}$ if $L$ odd), where we notice that $\delta_{\mathbb{Z}_2, \mathbb{Z}_1}$ is independent of $L$ to first order. Thus, for the $L=9$ chain we must have $\epsilon = \frac{1}{\sqrt{M}} \lesssim 0.058$, meaning $M = 500$ experimental shots would be sufficient to get an error $\epsilon \approx 0.044$ small enough to ensure distinguishability of the two quadratic SP decays at $T_{\text{PXP} | \mathbb{Z}_1} =  0.11 \mu\text{s}$. In practice, this number is likely to be an upper bound, and even less shots should be required to ensure distinguishability at later quartic timescales.

\section{Conclusion \label{sec:conclusion}}

We now summarize the results presented in this paper. We first showed that approximate quantum many-body scarring in the Rydberg chain, as modeled by the PXP model, gives rise to early-time SP decay with a characteristic non-generic rate determined solely by the overlap of the initial state with scarred eigenstates. The early-time SP decay shows a characteristic non-generic rate determined solely by the overlap of the initial state with scarred eigenstates. By deforming the PXP model with a revival-enhancing deformation, we show that this fact remains true regardless of the deformation strength within the enhancing range. In contrast, when an ergodicity-restoring deformation is applied, the SP decay rate displays distinct scaling compared to the case where QMBS are dominant. Additionally, we proved that this fact holds for QMBS models of the SGA class, where the early-time decay of perfectly scarred initial states is fully controlled by the QMBS sector and diverges from the decay of generic initial states which is controlled by the thermal sector. The SP thus provides, at very early times, information on QMBSs that is typically associated with late-time observables. We then constructed a simple model for LDOS of initial states having overlaps with both scar and thermal eigenstates, and quantified the minimal ratio of scar and thermal distribution variances for which scars fully control the early-time behavior of the survival probability, in the case where scar and thermal contributions to the LDOS are additive. Our results highlight the early-time SP decay rate as an easily accessible signature of slow thermalization and weak ergodicity breaking in QMBS systems. Recent works suggest that the PXP model is an ideal platform to test our proposal since its scarring properties are robust to finite temperature effects \cite{Desaules2023} and disorder \cite{MondragonShem2021}.

\begin{acknowledgments}
M.S. warmly acknowledges enlightening discussions with Daniel K. Mark. This work was supported by the Minist\`{e}re de l'\'{E}conomie, de l'Innovation et de l'Énergie du Qu\'{e}bec through its Research Chair in Quantum Computing, an NSERC Discovery grant, and the Canada First Research Excellence Fund. This work made use of the compute infrastructure of Calcul Québec and the Digital Research Alliance of Canada.
\end{acknowledgments}

\bibliography{biblio}

\end{document}